\documentclass[sigconf,screen]{acmart}

\usepackage{natbib}
\usepackage{url}
\usepackage{graphicx}
\usepackage{verbatimbox}
\usepackage[dvipsnames,table,xcdraw]{xcolor}
\usepackage{array}
\usepackage{multirow}
\usepackage[normalem]{ulem}
\useunder{\uline}{\ul}{}
\usepackage{booktabs}
\usepackage{supertabular,booktabs}
\usepackage{longtable}
\usepackage{lscape}
\usepackage{url}
\usepackage{dirtytalk}
\usepackage{graphicx}
\usepackage{comment}
\usepackage{makecell}
\usepackage{svg}
\usepackage{amsmath}
\usepackage{tabularx}
\usepackage{enumitem}
\usepackage{float}

\usepackage{tcolorbox}
\tcbuselibrary{skins}

\newenvironment{summarybox}{\begin{tcolorbox}[enhanced,arc=0mm,colback=gray!10,frame hidden,overlay unbroken={\draw[thick,black] (interior.north west)--(interior.south west);},left=2pt,right=0pt,top=0pt,bottom=0pt,before={\vspace{3pt}\noindent},after={\vspace{0pt}}]}
{\end{tcolorbox}}
\usepackage{tikz}

\usepackage[caption=false]{subfig}
\graphicspath{{Images/}}
\usepackage[normalem]{ulem}
\useunder{\uline}{\ul}{}

\usepackage{tabularx}
\usepackage{balance}

\AtBeginDocument{%
  }

\setcopyright{acmlicensed}
\copyrightyear{2018}
\acmYear{2018}
\acmDOI{XXXXXXX.XXXXXXX}
\acmBooktitle{Companion Proceedings of the 34th ACM Symposium on the Foundations of Software Engineering (FSE ’26), July 5 - 9, 2026 Montreal, Canada}
\acmISBN{978-1-4503-XXXX-X/2018/06}

\begin{document}

\title{Sustainable AI Assistance Through Digital Sobriety}


\author{Madeline Jennings}
\email{madeline.jennings@ucalgary.ca}
\affiliation{%
  \institution{University of Calgary}
  \city{Calgary}
  \state{Alberta}
  \country{Canada}
}

\author{Novarun Deb}
\email{novarun.deb@ucalgary.ca}
\affiliation{%
  \institution{University of Calgary}
  \city{Calgary}
  \state{Alberta}
  \country{Canada}
}

\author{Ronnie de Souza Santos}
\email{ronnie.desouzasantos@ucalgary.ca}
\affiliation{%
  \institution{University of Calgary}
  \city{Calgary}
  \state{Alberta}
  \country{Canada}
  }

\begin{abstract}
As AI assistants become commonplace in daily life, the demand for solutions that reduce the cost of inference without sacrificing utility is increasing. Existing work on AI sustainability frequently emphasizes hardware and software optimizations; however, there may be comparable value in social approaches that shape user behavior and discourage unnecessary use. In this study, we operationalize sustainability in terms of energy-efficiency and analyze a publicly sourced sample of prompts where AI is used for assistance in software development. Using this categorization, we find that nearly half of the observed queries can be considered unnecessary relative to their expected benefit. We further observe that factoid-style information retrieval constitutes the largest share of unnecessary requests, suggesting that a meaningful portion of everyday AI usage may be replaceable with lower-cost alternatives (e.g., conventional search or local documentation). These findings motivate a closer examination of how, why, and when AI systems are invoked, and what norms or interface-level nudges might reduce avoidable demand. We conclude with a call to replicate and extend this preliminary analysis and to pay greater attention to the social dimension of AI sustainability.
\end{abstract}




\maketitle

\section{Introduction}
\label{sec:introduction}

Rapid advances in machine learning technologies have allowed generative AI tools (Gen AI) to be accessible to everyday users at little to no cost. The deep-thinking capabilities of AI are often utilized in software contexts to create new code, review existing code, and answer questions about code; in 2025, 84\% of all developers surveyed use AI, and even 73\% of inexperienced developers use the technology \cite{stackoverflow2025Stack}. It is clear that experienced and inexperienced developers are using artificial intelligence to assist in learning \cite{keuning_students_2024} and increase productivity \cite{pereira_exploring_2025}. These advantages, however, come at a price: producing a single response involves billions of computations across billions of parameters, demanding enormous amounts of energy and water to operate at scale and satisfy widespread usage. The actual impact is difficult to quantify, and attempts to publicize sustainability metrics are often met with criticism claiming the data are intentionally misleading, as was the case with Google's 2025 AI report \cite{elsworth_measuring_nodate} \cite{pcgamerTheyreJust}. 

These concerns are not limited to model training. Inference at scale contributes significantly to energy consumption and carbon emissions \cite{vartziotis2024carbon, wilkins2024hybrid}. Current estimates of large language model (LLM) deployments indicate that operational inference workloads can consume substantial amounts of electricity when aggregated across millions of users \cite{vartziotis2024carbon}. Research on heterogeneous data center architectures demonstrates that even moderate improvements in task scheduling can reduce inference energy consumption, suggesting that LLM workloads are energy-intensive enough to warrant infrastructure-level optimization \cite{wilkins2024hybrid}. At the same time, empirical evaluations of LLM-generated code show that energy efficiency varies across programming languages and task types, reinforcing that the sustainability impact of AI systems extends beyond training and includes runtime behavior and deployment contexts \cite{solovyeva2025ai}. 

Overall, this body of work highlights that sustainability challenges arise not only from how models are built and deployed, but also from how they are used. In software engineering, sustainability is typically defined through efficiency metrics that seek to reduce unnecessary resource consumption \cite{calero_systematic_2013}. More broadly, sustainability has been framed as a multidimensional quality attribute encompassing environmental, economic, social, and technical concerns \cite{lago_framing_2015}. Within this perspective, reducing avoidable computation becomes a central design principle. By extension, one route to sustainable AI is to limit superfluous use of LLM systems.

In this context, digital sobriety, which is defined as using software only when it is genuinely necessary \cite{perea_digital_2023} \cite{clear_digital_2025}, offers a demand-side perspective on sustainable AI. While generative AI tools are frequently promoted as universal, all-purpose solutions, their unique strength lies in performing deep, context-sensitive reasoning. However, in practice, LLMs are often used as replacements for conventional search engines or documentation tools to answer trivial informational queries. If inference is energy-intensive at scale \cite{vartziotis2024carbon, wilkins2024hybrid}, then invoking such systems for tasks that do not require their reasoning capabilities may represent avoidable environmental cost.

The broader literature on Sustainable AI reinforces this distinction. Sustainable AI studies differentiate between \emph{AI for sustainability} and the \emph{sustainability of AI} itself \cite{van2021sustainable}. While AI for sustainability focuses on leveraging AI systems to achieve environmental goals, sustainability of AI emphasizes reducing the environmental footprint of AI systems across their lifecycle, including training, deployment, and governance \cite{van2021sustainable}. A systematic review of Green AI research shows that most existing work concentrates on monitoring model footprints, hyperparameter tuning, and benchmarking approaches, with comparatively limited attention to user behavior and demand-side mitigation strategies \cite{verdecchia_systematic_2023}. This suggests a gap in understanding how everyday usage patterns contribute to aggregate energy consumption.

This gap is particularly pronounced in software development contexts, where LLMs are deeply integrated into code editors and development workflows \cite{solovyeva2025ai}. While prior studies explore the energy efficiency of generated code \cite{solovyeva2025ai, vartziotis2024carbon}, fewer investigate whether invoking an LLM is necessary for a given task in the first place. From a sustainability standpoint, distinguishing between tasks that require deep reasoning and those that could be satisfied through lower-cost alternatives aligns with longstanding principles of need-based development and avoidance of unnecessary computation in green software engineering \cite{berleur_model_2010, lago_framing_2015}.

In this study, our objective is to quantify the potential impact of digital sobriety by scientifically categorizing user AI queries in the context of software. By distinguishing between queries that require deep, context-sensitive reasoning and those that could plausibly be addressed through lower-cost alternatives, we provide speculative impact estimates of superfluous LLM usage. This novel contribution complements infrastructure-level optimization efforts \cite{wilkins2024hybrid} and extends Green AI research beyond training-centric efficiency improvements \cite{verdecchia_systematic_2023}. Establishing an evidence-based baseline for unnecessary LLM usage is therefore essential for informing both research and practice on sustainable AI deployment \cite{van2021sustainable}.
 
From this introduction, this paper is organized as follows. Section 2 discusses industry perception of software sustainability and current approaches to green software; Section 3 outlines the rigorous manual study performed on the chosen AI query dataset and describes the classification schemes used; Section 4 summarizes the findings of the study through synthesized charts; Section 5 interprets the results and highlights their implications for research and practice.

\section{Background}
\label{sec:background}

As GenAI companies race to make the next technological breakthrough, researchers have been finding ways to provide incentives to these companies to adopt sustainable practices. Green software engineering initiatives outline sustainability as a property of quality software with economic, environmental and social dimensions \cite{calero_green_2015} \cite{lago_framing_2015}, although it is unfortunately common for these to be ignored in favor of more direct and immediate benefits. 

Within this framing, sustainability requires attention not only to technological performance but also to patterns of resource use across the software lifecycle \cite{berleur_model_2010, lago_framing_2015}. Sustainable AI scholarship further distinguishes between leveraging AI to support sustainability goals and reducing the environmental footprint of AI systems themselves \cite{van2021sustainable}. This distinction is important because the rapid diffusion of generative AI tools into everyday life means that their aggregate environmental impact depends not only on model design but also on how frequently and for what purposes they are used.

The primary sustainability focus of AI researchers and practitioners lies in Green IT, i.e., reducing the consumption of energy and water in performing software tasks \cite{berleur_model_2010}. Current approaches to Green IT in AI generally belong to one of two categories: improvements to AI models themselves, through fine-tuning or algorithm design to reduce the resource requirements of a task; and increased awareness of environmental factors, through understanding of energy-precision trade-offs and live footprint monitoring \cite{verdecchia_systematic_2023}. The existing understanding of software sustainability is often fragmented; practitioners primarily quantify the environmental impact of applications through only their Scope 2 emissions, defined by the Greenhouse Gas Protocol as the emissions required to produce the energy consumed \cite{mcmahon_project_nodate}. Existing estimates frequently understate power consumption because it is difficult to account for indirect energy uses, such as the energy required to source the water that cools the hardware \cite{caravaca_prompts_2025} \cite{elsworth_measuring_nodate}.

Although early Green AI discussions emphasized the carbon intensity of model training, more recent work shows that inference at scale represents a substantial and continuous source of energy demand \cite{vartziotis2024carbon, wilkins2024hybrid}. Operational inference workloads for widely deployed LLM systems can consume significant amounts of electricity when aggregated across millions of daily users \cite{vartziotis2024carbon}. Infrastructure-level studies demonstrate that workload-aware scheduling can reduce inference energy consumption, confirming that the act of serving queries itself carries measurable environmental cost \cite{wilkins2024hybrid}. These findings suggest that sustainability challenges persist throughout the deployment phase of AI systems, not only during model development.

Importantly, inference demand scales with everyday user behavior. As LLM-based assistants become embedded in web browsers, mobile applications, and search interfaces, individuals increasingly rely on them for information retrieval, summarization, drafting, and general knowledge queries. From a sustainability perspective, replacing traditional search engine queries or static documentation lookup with generative model invocations may increase per-query computational intensity. Since inference energy accumulates across millions of interactions \cite{vartziotis2024carbon}, even marginal increases in average query cost can translate into substantial aggregate environmental impact.

Research efforts in developing sustainable AI solutions have approached power consumption reduction tasks using various approaches. One prominent approach to Green AI is in carbon-aware scheduling tools for cloud-computing software: considering the physical location of servers can affect not just energy costs but Scope 2 emissions as well by directing requests to low-carbon power grids in both training of models and their actual service \cite{souza_casper_2024} \cite{chien_reducing_2023} \cite{iftikhar_enhancing_2025}. While such infrastructure-level optimizations can reduce the carbon intensity of computation, they do not directly address the volume of inference requests generated by widespread adoption.

A systematic review of Green AI research notes that most existing work concentrates on model-level optimization and benchmarking, with comparatively limited attention to behavioral or demand-driven mitigation strategies \cite{verdecchia_systematic_2023}. This omission is particularly relevant in everyday contexts, where LLMs are often used as general-purpose tools for tasks that may not require deep, context-sensitive reasoning. If inference is energy-intensive at scale \cite{vartziotis2024carbon, wilkins2024hybrid}, invoking generative AI for trivial informational queries may represent avoidable resource consumption.

There is currently almost no focus on digital sobriety as a strategy for achieving Green AI, which provides the main motivation for this study. Digital sobriety shifts attention from optimizing how efficiently a query is processed to questioning whether the query should be processed by an LLM at all. In alignment with need-based development principles in green software engineering \cite{berleur_model_2010, lago_framing_2015}, digital sobriety frames unnecessary inference as avoidable resource consumption. Given that inference workloads contribute materially to AI’s operational footprint \cite{vartziotis2024carbon, wilkins2024hybrid}, understanding patterns of everyday LLM usage becomes essential for developing demand-aware sustainable AI practices.

\section{Method}
\label{sec:method}

This study is defined as an empirical analysis of user prompts drawn from a large-scale conversational dataset to estimate the prevalence of unnecessary generative AI use in software assistance contexts. We apply systematic filtering criteria to isolate software-related queries, then manually classify a random sample according to predefined categories of uses and reasoning complexity. By operationalizing necessity in terms of whether a task requires complex, context-sensitive reasoning that justifies AI inference, we quantify the proportion of queries that could plausibly be satisfied through lower-cost alternatives. This procedure enables an evidence-based assessment of the potential contribution of digital sobriety to Green AI. The following research questions guided the design of this study:
\newcommand\rqone{Is there a considerable amount of unnecessary AI use?}
\newcommand\rqtwo{Do correlations exist between AI use cases and unnecessary AI usage?}
\newcommand\rqthree{Can we conclude that digital sobriety would benefit Green AI?}

\begin{description}
    \item[\textit{RQ1}] \rqone
    \item[\textit{RQ2}] \rqtwo
    \item[\textit{RQ3}] \rqthree
\end{description}

\subsection{Synthesis of Definitions}
To address RQ1, we first define what it means for AI use to be sustainable in terms of whether an AI tool is \,necessary\, to complete a user’s task. Because prior work offers few rigorous methods for labeling queries as necessary or not, we synthesize operational definitions to fill this gap. Here, necessity is evaluated through resource use: when a non-AI alternative can deliver comparable assistance while consuming fewer resources, the AI-mediated query is treated as unnecessary. Accordingly, an \emph{unnecessary query} is (i) a question-answering task that can be answered adequately without AI, or (ii) a generation task that can be satisfied using readily available existing content or knowledge---assuming that, for context-independent tasks, non-AI tools can achieve similar outcomes at lower cost. Consider the case where a simple Google search provides a solution to a user's query; the use of AI is unnecessary and wasteful in this context.

To systematically determine whether a query genuinely benefits from AI, we consider the kinds of higher-order reasoning that AI systems can provide. Drawing on the Researchy Questions dataset \cite{rosset_researchy_2024}, we characterize \,complex\, queries as those requiring ``slow thinking''\cite{chalmers}, including contextual interpretation and decision-making. By contrast, queries composed only of ``factoid'' sub-tasks\cite{rosset_researchy_2024} involve little to no deeper reasoning and are undesirable in the context of AI assistance. That said, some requests may not demand deep thought yet still produce nontrivial value---for example, generating boilerplate code for a standard solution while adapting names, interfaces, or constraints. Because the necessity of such convenience-oriented tasks is inherently subjective, we treat them as a separate category. To distinguish convenience tasks from complex tasks in a reproducible way, we require that explicit criteria be met for a task to qualify as complex; the resulting category definitions are summarized in Table \ref{tab:QCmplxCat}.

For RQ2, we adopt use-case definitions from the National Institute of Standards and Technology (NIST) AI Use Taxonomy \cite{theofanos_ai_2024}, from which a subset of human-AI activities were customized to fit the software focus of the study. The use cases included in this study and their definitions are listed in Table \ref{tab:AIUseCase}. \begin{center}
\begin{table}
\caption{Query Complexity Categories 
\label{tab:QCmplxCat}\cite{rosset_researchy_2024}}
\begin{tabular}{ | m{2cm} | m{5.4cm} |} 
\hline
\small \textbf{Category} & \small \textbf{Definition} \\ 
\hline
 \small Factoid Query & \small A question-answering task with a correct and/or canonical answer; or a generation task that can be completely fulfilled with a common existing solution; an unnecessary AI task. \\ 
 \hline
 \small Complex Query & \small A non-factoid task that requires analytical thinking, via the evaluation of contexts and/or decision making; OR relational thinking, via utilization of multiple, combined contexts to perform a task; OR deep thinking, via the synthesis of factoid contexts into non-factoid contexts.\\  
 \hline
 \small Convenience Query & \small A non-factoid task that fails to demonstrate complex thinking, usually for incremental productivity benefits; tasks with subjective necessity.\\
 \hline
\end{tabular}
\end{table}
\end{center} \begin{center}
\begin{table}
\caption{AI Assistance Use Cases 
\label{tab:AIUseCase}\cite{theofanos_ai_2024}}
\begin{tabular}{ | m{2cm} | m{5.4cm} |} 
\hline
\small \textbf{Use Case} & \small \textbf{Description} \\ 
\hline
\small Content Creation & \small Generating text or code from a provided prompt or description \\ 
\hline
\small Content Synthesis & \small Modifying, translating, paraphrasing, or retrieving information from provided text or code\\  
\hline
\small Information Retrieval & \small Finding answers to questions, providing facts, or acting as a search engine\\
\hline
\small Decision Making & \small Selecting from a set of possible alternatives, including yes / no\\
\hline
\small Process Automation & \small Answering math or logic problems, performing arbitrary or repetitive tasks\\
\hline
\small Recommendation & \small Suggesting from a set of viable options\\ 
\hline
\small Problem Detection & \small Identifying the cause of problems\\
\hline
\end{tabular}
\end{table}
\end{center}
\vspace{-0.9cm}
\subsection{Data Selection Process}
The data set chosen for the procedure was sourced from LMSYS-Chat-1M, a publicly sourced data set of user queries across multiple AI providers and use contexts \cite{2309.11998}. Only the first prompt field, i.e. \verb|entry['conversation'][0]['content']|, is considered and used for the experiment. After retrieving the data set using Hugging Face libraries in Python, filters were applied to case-insensitive prompts (refer to Fig. \ref{fig:FilterCond}) to meet the exclusion criteria outlined in Table \ref{tab:ExclCrit}. From the initial 1,000,000 queries, 987,560 were filtered to obtain 12,440 potential candidates. Of these candidates, 200 were randomly selected to obtain the final experimental data set.

\begin{figure}
\begin{center}
\begin{tcolorbox}
\begin{verbbox}
    CONTAINS
        ["c++","c#","java","javascript","php","css","html",
        "typescript","computer","software","program",
        "application","algorithm"]
    AND REGEX 
        (?:^|[.!?;:,]\s*)(how|who|what|where|when|
        which|why|will|explain|elaborate)\b
    AND NOT CONTAINS 
        ["role play","roleplay","offensive","inappropriate",
        "fictional","play the role","jailbreak"]
\end{verbbox}
\resizebox{1\textwidth}{!}{\theverbbox}
\end{tcolorbox}
\caption{Initial Filter Conditions}
\label{fig:FilterCond}
\end{center}
\end{figure}

\begin{table}
\begin{center}
\caption{Data-Selection Exclusion Criteria}
\label{tab:ExclCrit}
\begin{tabular}{ | m{1cm} | m{4.4cm} | m{1.4cm} |} 
    \hline
    \thead{ID} & \thead{Description} & \thead{Count} \\ \hline
    EXC1&Prompt was not marked as English&222547\\ \hline
    EXC2&Prompt exceeded 2048 characters in length&21153\\ \hline
    EXC3&Prompt did not start a sentence with a prompting keyword&525068\\ \hline
    EXC4&Prompt did not contain a software-related keyword&217872\\ \hline
    EXC5&Prompt included a blacklisted phrase&920\\
    \hline
\end{tabular}
\end{center}
\end{table}

\subsection{Experimental Procedure}
While prior work labels large volumes of user queries using automated LLM-based classifiers\cite{jo_taxonomy_nodate}, those taxonomies rely on more concrete criteria than the use case considered in this study. Because LLMs function as black boxes and their decisions are not easily audited, the queries must be manually reviewed and coded by a human researcher to preserve explainability and transparency in the experimental procedure, assigning both AI use case and prompt complexity. The definitions in Table \ref{tab:QCmplxCat} serve as the coding criteria for thought complexity: prompts with a correct or otherwise known answer are classified as factoid; if any task (or required sub-task) satisfies one of the three complex-thinking criteria, the prompt is classified as complex; if neither applies, or if the fit is unclear, it is classified as convenience. Figure \ref{fig:ExpProc} summarizes the steps of this procedure.

\begin{figure}[b]
\begin{center}
\fbox{\begin{minipage}{22em} 
\small 1. SELECT data sample of user queries

\small 2. FOR each query, IF exclusion criteria is NOT met:

\begin{itemize}
\small \item[a.] ASSIGN query use case
\small \item[b.] ANALYZE query for complex thought requirements
\small \item[c.] ASSIGN query complexity based on definitions
\end{itemize}

\end{minipage}}
\caption{Experimental Procedure Summary}
\label{fig:ExpProc}
\end{center}
\end{figure}

\begin{table}
\begin{center}
\renewcommand{\arraystretch}{1.5}
\caption{In-Procedure Exclusion Criteria}
\label{tab:ExclCrit2}
\begin{tabular}{ | m{1cm} | m{4.4cm} | m{1.4cm} |} 
    \hline
    \thead{ID} & \thead{Description} & \thead{Count} \\ \hline
    EXC6&Prompt engaged in roleplay scenarios outside of prompt engineering, including "jailbreak" attempts&6 \\ \hline
    EXC7&Prompt can not be interpreted, i.e. poor english&5 \\ \hline
    EXC8&Prompt contains obscene, explicit, or harmful material&4 \\ \hline
    EXC9&Prompt does not ask for any type of AI assistance; a statement&10 \\ \hline
    EXC10&Prompt is a repeat and has already been analyzed; may have different context&29 \\ \hline
\end{tabular}
\end{center}
\end{table}

\section{Findings} \label{sec:findings}

To answer the proposed research questions, each of the 3 reviewers individually performed a manual analysis of the selected data set and then consolidated to obtain a final set of results. For each of the 200 queries selected for analysis, each reviewer first determined if any exclusion criteria were met. If the criteria did not meet any of the exclusion criteria, they assigned values describing the prompt's most relevant AI assistance use case (see Table \ref{tab:AIUseCase}) and the level of complexity required to provide an answer (see Table \ref{tab:QCmplxCat}). To mitigate potential threats to validity, reviewer 1 analyzed all 200 tuples, reviewer 2 analyzed the first 100, and reviewer 3 analyzed the last 100 to be combined into a complete set of 200. The two sets were then cross-compared for conflicting assignments and meetings were held to resolve disputes in assignments until a final, unanimous decision was agreed upon. A summary of the main findings including use case breakdown, query complexity by use case, and summarized totals can be found in Table \ref{tab:ResSum} and are visualized in Figure 3.

Of the 200 queries randomly selected from the dataset, 52 additional queries were manually filtered during data evaluation according to exclusion criteria 6--10 (listed in Table \ref{tab:ExclCrit2}), producing a final set of 148 queries for analysis. Because the initial exclusion criteria (Table \ref{tab:ExclCrit}) were intentionally lightweight, the remaining undesirable prompts were removed by an additional manual screening step. The extracted data and the procedure replication package are provided in Section \ref{sec:sec7}.

Of the 148 queries, 64 were determined to be factoid queries - just under half of the considered sample at 43.2\%. We consider this sample in the context of sustainability discussed in Section 3: these queries did not require energy-intensive AI systems to answer, and therefore resources were inefficiently utilized by providing an AI inference. 
Common examples of factoid queries obtained from the sample include code documentation retrieval ("How do I do this in Python?"), search queries ("How does merge sorting work?"), finding information within provided context ("How many times is the phrase mentioned here?"), and generation of canonical texts ("Write a 'hello world' function in C++"). 

Of the remaining queries, 61 demonstrated complex thought through analytical, relational, and/or deep-thinking reasoning. The key difference between factoid and complex queries lies in the utilization of user-provided contexts; complex queries require natural language processing to fulfill. These often come in the form of recommendation and evaluation tasks ("what approach should I use") and summarization tasks ("summarize the following text"). It is evident that complex thought requirements show a strong correlation with use cases of decision making, recommendation, and content synthesis, which aligns with expectations given the complex thought criteria. However, complexity is not exclusively limited to these use cases: a statistically significant distribution of use cases were detected across considered use cases (not including problem detection and process automation due to lack of data). 

Of the 23 convenience queries, there is an evident trend towards content creation tasks, making up 11 of the 18 recorded under the use case category. These samples primarily consisted of generation tasks that are non-determinate in their output but logically could be derived from existing factoids; for instance, asking for the generation of an existing function with different variable names produces a non-factoid response but with certain keywords replaced. It is indeterminate whether the replacement of keywords requires complex reasoning through the AI's understanding of the code or if it is performing a simple word-replace. Thus, the necessity of convenience cases in a sustainability context would require further information to determine. These convenience cases are therefore excluded from the general consideration of this study and from the answer to RQ1 and RQ2.

\begin{table}[H]
\begin{center}
\renewcommand{\arraystretch}{1.3}
\caption{Summarized Results of Experimental Procedure}
\label{tab:ResSum}
\resizebox{0.49\textwidth}{!}{%
\begin{tabular}{ |m{1.4cm}|m{0.7cm}|m{0.7cm}|m{0.7cm}|m{0.7cm}|m{0.7cm}|m{0.7cm}|m{0.7cm}| }
\hline
\thead{Use Case} & \thead{Total\#} & \thead{Fact\#} & \thead{Conv\#} & \thead{Comp\#} & \thead{Fact\%} & \thead{Conv\%} & \thead{Comp\%} \\ \hline
    Content Creation & 18 &0 & 11 & 7 & 0.0\% & 61.1\% & 38.9\% \\ \hline
    Content Synthesis & 26 &2 & 4 & 20 & 7.7\% & 15.4\% & 76.9\% \\ \hline
    Information Retrieval & 75 & 57 & 5 & 13 & 76.0\% & 6.7\% & 17.3\% \\ \hline
    Decision Making & 11 & 2 & 0 & 9 & 18.2\% & 0.0\% & 81.8\% \\ \hline
    Process Automation & 2 & 1 & 0 & 1 & 50.0\% & 0.0\% & 50.0\% \\ \hline
    Recommen
    -dation & 14 & 1 & 3 & 10 & 7.1\% & 21.4\% & 71.4\% \\ \hline
    Problem Detection & 2 & 1 & 0 & 1 & 50.0\% & 0.0\% & 50.0\% \\ \hline
    Total&148&64&23&61&43.2\%&15.5\%&41.2\%\\ \hline
\end{tabular}%
}
\end{center}
\end{table}

\begin{figure}[t]
\begin{center}
\includegraphics[width=0.5\textwidth]{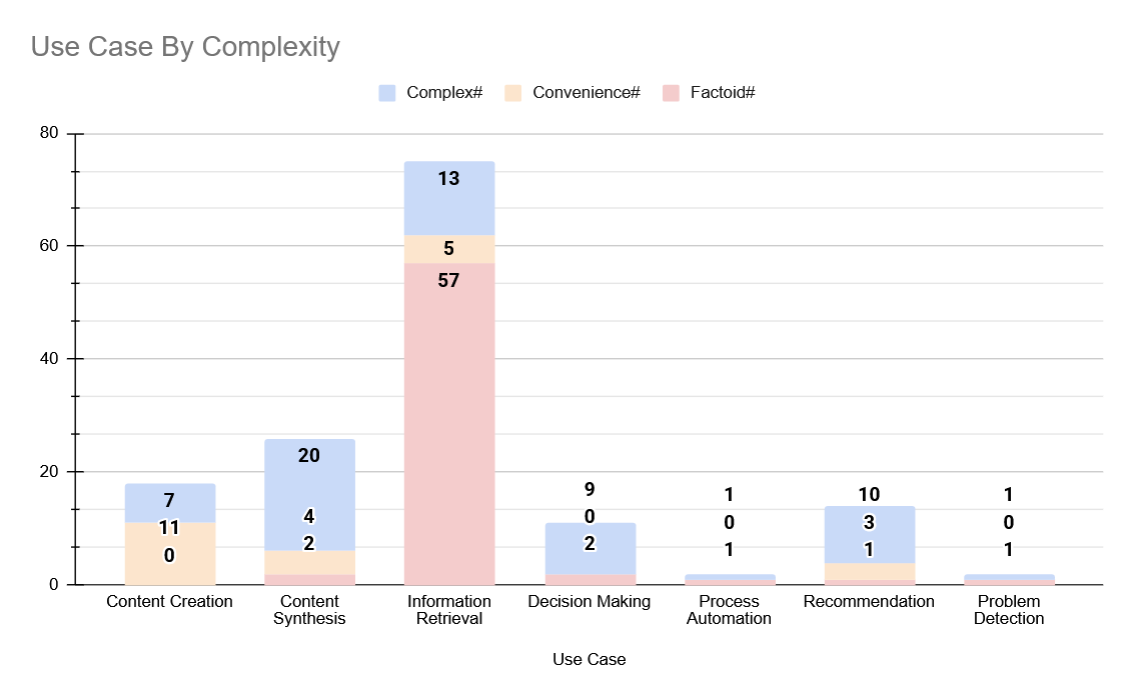}
\caption{Use Case By Complexity Chart}
\end{center}
\end{figure}

\begin{summarybox} 
RQ1) \rqone

\small \emph{Yes, studies performed on the dataset show a surprisingly large quantity of unnecessary (i.e. unsustainable) AI usage at 43.2\%.}
\end{summarybox}

Only 2 process automation and problem detection cases were identified: due to their small sample size, their statistics will not be considered in answering RQ2. Of the 75 queries that tasked the AI with information retrieval, an overwhelming 76\% were factoids. Factoid queries were identified for each use case, but no more than 3 were identified for a single category other than information retrieval. This implies that while there is an unnecessary use of AI in assistance contexts, there is a very significant correlation between AI for information retrieval and unnecessary AI use. This is not to discount the remaining percentage of non-factoid assignments\%: it is clear that information retrieval is not exclusively a factoid task and suggests deeper trends in the ways people are using AI to retrieve information.

Consider two similar prompts where one asks the AI to find any text editor and one that asks the AI to find the best text editing application for software development. No complex thinking is required to retrieve a potential candidate application without further criteria; we do not know how the AI is making its decision, and as such its results are no different than a google search to retrieve a text editor. However, if the prompt instead asks the AI to identify the "best" application for software development, it must engage in analytical and relational reasoning to compare and assess possible options against criteria it has to infer as being most appropriate for software development. Existing tools may be able to provide similar answers, i.e. articles made by humans, but are evidently not capable of human-level thinking. This becomes increasingly apparent as prompt complexity increases - consider the previous example but with specific evaluation criteria suited to the user's personal needs.

As one would assume, tasks based on analytical reasoning are more likely to utilize complex AI thought: 76.9\% of content synthesis queries, 81.8\% of decision making queries, and 71.4\% of recommendation queries were complex and therefore used AI sustainably by the chosen criteria. The content creation queries analyzed were primarily indeterminate in terms of necessity, with 61.1\% categorized as convenience queries, implying that the generation of text and code is used primarily without contextual thinking previously-discussed. Therefore, the sustainability of content creation AI queries should be evaluated through a cost-benefit comparison between the energy consumed and the other resources they save, such as time and labor.

\begin{summarybox} 
RQ2) \rqtwo

\small \emph{Most information-retrieval queries were unnecessary and thus unsustainable. In contrast, queries involving content synthesis, decision-making, and recommendations were generally sustainable because they genuinely needed AI support. Content creation showed mixed sustainability and needs to be evaluated individually. These trends were not absolute, with both sustainable and unsustainable exceptions in every category.}
\end{summarybox}

By establishing clear context-independent criteria that can be applied to any prompt, this study suggests that prompt-level sustainability can be empirically assessed. Although manual analysis is too limited to claim this conclusively, the results still indicate potentially important social patterns that may be driving unsustainable AI use and warrant further investigation. If these findings generalize beyond the sample, then understanding how people use AI becomes a key lever for both environmental and economic gains. Under optimistic assumptions—using estimates of roughly 0.3\,Wh per AI query versus 0.03\,Wh for a search-engine query \cite{elsworth_measuring_nodate} \cite{holzle_powering_nodate}—reducing AI use by 90\% in about half of cases would imply an overall energy reduction of about 45\%. Given the scale of AI deployment, even capturing a fraction of these improvements could yield meaningful savings.

\begin{summarybox} 
RQ3) \rqthree

\small \emph{Yes; assuming the findings of the study hold true and using existing estimates, it then follows that digital sobriety could reduce overall energy usage by up to 43\%. This can be further increased to an estimated 58.7\% if convenience queries are considered unsustainable and therefore unnecessary. Further studies on digital sobriety in Green AI are required to solidify these findings.}
\end{summarybox}

\section{Discussion} \label{sec:discussion}
In this section, we compare our results with the general body of work on sustainability and AI identified in the literature. Additionally, we discuss the implications of our study as well as methodological threats to validity.

\subsection{Comparing Results and Novelty}

As described in Section 2, prior Green AI research primarily approaches sustainability through improvements in model efficiency and infrastructure optimization \cite{verdecchia_systematic_2023, wilkins2024hybrid}. These studies demonstrate that carbon intensity per inference can be reduced through technical interventions such as workload-aware scheduling and carbon-aware routing \cite{souza_casper_2024, wilkins2024hybrid}. Our results extend this line of work by shifting attention from how efficiently inference is served to how frequently it is invoked. While existing research treats demand as given, our findings show that a substantial proportion of LLM queries in software-related contexts do not require deep, context-sensitive reasoning, with unnecessary usage concentrated in information retrieval tasks.

Sustainable AI studies emphasize lifecycle accountability and call for reducing the environmental footprint of AI systems across development and deployment phases \cite{van2021sustainable}. However, systematic reviews indicate limited empirical investigation of demand-side mitigation strategies \cite{verdecchia_systematic_2023}. Our study contributes to this topic by providing an empirical estimate of potentially avoidable LLM usage and modeling its associated energy implications. By quantifying factoid queries and exploring their substitution potential, we introduce behavioral demand as a measurable component of AI sustainability.

Finally, green software engineering literature has long advocated avoiding unnecessary computation as a sustainability principle \cite{berleur_model_2010, lago_framing_2015}. Our contribution operationalizes this principle in the context of generative AI. The novelty of our study lies in demonstrating that digital sobriety can function as a concrete sustainability strategy for AI and LLM systems by reducing avoidable inference demand. In a discourse largely centered on model optimization and infrastructure efficiency \cite{verdecchia_systematic_2023, wilkins2024hybrid}, our findings add a complementary behavioral dimension. By grounding digital sobriety in empirical usage patterns and linking it to measurable energy implications, we contribute to ongoing efforts to make AI systems more environmentally sustainable \cite{van2021sustainable}.

\subsection{Implications to Research}
This study provides motivation to consider digital sobriety in Green AI by providing significant findings within its limited scope. Future research of the methodology and findings through cross-validation between different samples or data sets is required to solidify the sustainability trends revealed by this study. Further studies on social sustainability factors, in particular how and when digital sobriety is applicable to AI, would prove beneficial to Green AI. The continuation of this study's preliminary findings through the development of rigorous complexity categorization procedures, including the potential training and use of LLMs to discern a prompt's complexity, is called for.

\subsection{Implications to Practice}
This study proposes that, even without regulatory oversight, there are potentially great financial incentives in Green AI through digital sobriety. Current AI systems are marketed as general-purpose assistants, which this study may demonstrate leads to the inefficient allocation of increasingly-powerful AI where it is not necessary to provide answers of similar quality. An approach AI providers could consider would be software architectures that allocate LLM resources based on demand, allowing equivalent-quality answers with lower resource expenditure and therefore cost savings and emission reductions. Practitioners may also consider how the marketing of their applications as general-purpose agents rather than heavy-duty tools contributes to unsustainable use. Offering lower-complexity models and encouraging user mindfulness of the cost of inference through pay-per-use models puts the onus of sustainability onto the user. Both potential approaches imply an economic niche for small language models, implying that attention should be brought away from bigger, more accurate models and towards smaller, more efficient ones. 

\subsection{Threats to Validity}
Given the limited scope of the study and the novelty of the method, the main validity risk is that the patterns observed in this sample may not reflect broader trends in AI usage. A single dataset, LMSYS-Chat-1M \cite{2309.11998} is analyzed, from which 148 of 200 randomly-sampled queries are considered for analysis. Threats to validity present within the dataset itself, including biases stemming from the open-source collection methodology, thereby carry over to this study. The simple randomly-selected nature of the sample, in addition to previously discussed limitations of scope, restricts this study's findings from being generally applicable. Although the results of this study are explicitly framed as preliminary for these reasons, they could still be misleading. The manual coding approach, chosen to support replication and cross-validation, also introduces additional threats. In addition, the small sample size prevented any meaningful conclusions about sustainability in uncommon use cases, namely process automation and problem detection; larger datasets are needed to establish early trends for these categories and to reinforce the findings reported here. Finally, the results may be influenced by the bias of the reviewers in how the queries were categorized.
\section{Conclusion}
\label{sec:Concl}
The technical pathways to sustainable AI are widely studied, but existing work pays far less attention to the social drivers of sustainability. Framing sustainability around energy efficiency and treating prompts that do not use or need complex reasoning as inefficient, this paper offers preliminary evidence to support digital sobriety in AI use. Of 148 non-excluded user queries, at least 43.2\% were determined to be answerable via conventional search methods, suggesting that this pattern may extend beyond the sampled setting and across AI applications. Steering users away from AI for simple information requests while reserving it for tasks that genuinely benefit from complex reasoning could substantially lower energy use and associated emissions. These conclusions remain tentative and require further validation; however, if confirmed, they point toward a shift to smaller, lower-compute models when the task is simple.

\section*{Data Availability} \label{sec:sec7}

The spreadsheet containing the 12,440 candidate queries considered during data selection—together with the sample used and any data extracted from the experimental procedure—is available online.\footnote{\url{https://docs.google.com/spreadsheets/d/e/2PACX-1vQim-gwEx1r9pe6YCY_tLUP53sZi_6bkgEbK878zo1zmcU11q3TtsD06rnaKEeNo2iUWHUIOlI1pNzN/pubhtml}}
The Python code used to retrieve and filter tuples during data selection is publicly accessible.\footnote{\url{https://anonymous.4open.science/r/llm-digital-sobriety-retrieval-19C1/}} 

\nocite{*}

\bibliographystyle{ACM-Reference-Format}
\bibliography{bib.bib}

\appendix

\end{document}